\documentstyle[preprint,aps,eqsecnum]{revtex}
\begin{document}
\draft

%%%%%%%%%%%%%%%%%%%%%%%%%%%%%%%%%%%%%%%%%%%%%%%%%%%%%%%%%%%%%%%%%%%%%%%%%%%%%%%
%   Title page

\title{
\begin{flushright}
{\large IP-ASTP-05-95}
\end{flushright}
\Large\bf Light-Front Heavy Quark Effective Theory \\
 and Heavy Meson Bound States
}
%%%%%%%%%%%%%%%%%%%%%%%%%%%%%%%%%%%%%%%%%%%%%%%%%%%%%%%%%%%%%%%%%%%%%%%%%%%%%%%
\author{{\bf Chi-Yee Cheung, Wei-Min Zhang}\\
        Institute of Physics, Academia Sinica, Taipei 11529, Taiwan \\
                        and \\
               {\bf Guey-Lin Lin}\\
        Institute of Physics, National Chiao-Tung University, Hsinchu
                30050, Taiwan \\
}
%%%%%%%%%%%%%%%%%%%%%%%%%%%%%%%%%%%%%%%%%%%%%%%%%%%%%%%%%%%%%%%%%%%%%%%%%%%%%%%
\date{March 20, 1995}
\maketitle

%%%%%%%%%%%%%%%%%%%%%%%%%%%%%%%%%%%%%%%%%%%%%%%%%%%%%%%%%%%%%%%%%%%%%%%%%%%%%%%
\begin{abstract}
The heavy quark effective theory is developed on the light-front.
Based on this effective theory,
a light-front heavy meson bound state with
definite spin and parity is constructed.
Within the effective theory, the Isgur-Wise function is derived in
terms of the asymptotic light-front bound state amplitudes in
the limit $m_Q\rightarrow\infty$;
the result is a general expression for arbitrary recoil velocities.
With the asymptotic form of the BSW amplitudes,
the Isgur-Wise function is given by $\xi(v\cdot v')=1/v\cdot v'$.
The slope at the zero-recoil point is $\rho^2=-\xi'(1)=1$,
in excellent agreement with the recent CLEO result of
$\rho^2=1.01\pm0.15\pm0.09$.
\end{abstract}

\vspace{0.4in}

\pacs{PACS numbers: 11.10.Ef, 12.39.Hg, 14.40.Lb, 14.40.Nd}

%%%%%%%%%%%%%%%%%%%%%%%%%%%%%%%%%%%%%%%%%%%%%%%%%%%%%%%%%%%%%%%%%%%%%%%%%%%%%%%
\newpage
\baselineskip .29in

\section{Introduction}

The discovery of heavy quark symmetry (HQS) \cite{Isgur90}
and the subsequent construction of
the heavy quark effective theory (HQET) \cite{Hill,Georgi}
have led to intense activities
in the study of heavy hadron physics,
and much progress has been made in recent years \cite{review}.
The so called HQET is an effective theory of
quantum chromodynamics (QCD)
valid in situations where the gluon momenta ($\sim\Lambda_{QCD}$)
are much smaller than the heavy quark masses ($m_Q$).
In effect, the HQET provides us with a systematic expansion
of the QCD Lagrangian in terms of
the dimensionless parameter $\Lambda_{QCD}/m_Q$
\cite{Grinstein,Hill2,Luke,Falk,Mannel}.
In the symmetry limit ($m_Q\rightarrow\infty$),
the coupling between heavy quark and gluon becomes independent of the
spin and flavor of the heavy quark.  Thus the leading order effective
Lagrangian possesses a new spin-flavor symmetry,
which is not manifest in the original QCD Lagrangian.

Since HQS is a symmetry of QCD for heavy quarks at the confinement scale,
it can therefore be used to extract model independent dynamical consequences
of the theory at a scale where perturbative calculations are not possible.
In practical applications,
the HQS is most useful in reducing the number of independent
form factors in various heavy hadron decays, and thereby greatly
simplifying the complexity of theoretical analyses.
For instance, in the symmetry limit,
all of the form factors in $B \rightarrow D$ and $B \rightarrow D^*$ are
related by spin symmetry to a single universal function,
called the Isgur-Wise function.  Moreover,
the normalization of this universal function at the zero-recoil
point is also fixed by flavor symmetry, which then
permits a model-independent means of extracting the important
Kobayashi-Maskawa matrix element $|V_{cb}|$ from experimental data.
Similarly in heavy baryon decays,
such as $\Lambda_b \rightarrow \Lambda_c$ \cite{Isgur2},
the application of HQS also leads to tremendous simplifications.
Although HQS was first discovered in the weak decays of heavy hadrons,
it has since found applications in many other areas of heavy hadron
physics.  For example, by combining HQS with chiral symmetry
it is possible to construct a chiral Lagrangian
for the low energy interactions of heavy hadrons with
Goldstone bosons. \cite{Yan,Wise,BD}.  This theory has
been extended to include heavy-flavor-conserving weak decays \cite{CLY1},
as well as electromagnetic interactions \cite{CLY2,ChoG}.
Furthermore, HQS has also been applied in inclusive $B$ meson decays,
where the main thrust was to reliably extract the K-M matrix
element $V_{ub}$ from the end point spectrum of the charge lepton
\cite{Chay}.

Beyond the symmetry limit,
the HQET serves as a theoretical framework for the systematic
computation of $1/m_Q$ corrections.
However, in order to make definite predictions, it is also necessary
to construct explicitly the heavy hadron bound state wave functions
within the HQET.  This is of course a difficult task,
to which a satisfactory solution does not exist.
Nevertheless, we do expect that HQS will led to
considerable conceptual and calculational simplifications.
One of the purposes of this paper is to lay the ground work for
solving this important problem on the light-front.

In order to better motivate the work of
this paper as well as to be self-contained, we present
below a brief description of the HQET in the equal-time form,
and point out the issues to be addressed
in this paper as we proceed.
Let us start with the QCD Lagrangian for a heavy quark,
\begin{equation}
        {\cal L} = \bar{Q} ( i \not{\! \! D} - m_Q) Q ,\label{1.1}
\end{equation}
where $Q$ is the heavy quark field operator, $m_Q$ the heavy quark mass,
and $D^{\mu} = \partial^{\mu} - igT_a A^{\mu}_a$ the QCD covariant
derivative.  The pure gauge part of the QCD Lagrangian has not been
included because it is irrelevant for our discussions.
The HQET in the usual equal-time formalism is obtained simply by
redefining the heavy quark fields as follows:\cite{Georgi}:
\begin{equation}
        Q(x) = e^{-i m_Q v \cdot x} [h_v(x) +H_v(x)]   \label{rdhq1},
\end{equation}
where $v^{\mu}$ is the four velocity of the heavy hadrons $v^2=1$;
$h_v(x)$ and $H_v(x)$ are respectively the large and small
components of $Q(x)$, such that
\begin{eqnarray}
\not{\! v}h_v(x)&=&h_v(x)\nonumber\\
\not{\! v}H_v(x)&=&-H_v(x).           \label{I1}
\end{eqnarray}
This phase redefinition amounts to a splitting of the
heavy quark momentum: $p=m_qv +k$, where $k\sim\Lambda_{QCD}$
is called the residual momentum which measures the fluctuation
around the mass shell.
With such a redefinition, the heavy quark Dirac equation is reduced to
\begin{equation}
i\! \not{\! \! D} h_v + i(\not{\! \! D} -2m_Q) H_v = 0.   \label{I2}
\end{equation}
Which can be further decomposed into two coupled equations,
{\it viz}.,
\begin{eqnarray}
& & - i v \cdot D h_v = i\not{\! \! D}_{\bot} H_v\nonumber\\
& & (i v \cdot D + 2m_Q) H_v = i\not{\! \! D}_{\bot} h_v,
\end{eqnarray}
where $D_{\bot}^{\mu} \equiv D^{\mu} - v^{\mu} v \cdot D$.
Thus, one can express $H_v(x)$ in terms of $h_v(x)$ and
show that $H_v(x)$ is suppressed by $1/m_Q$ compared to $h_v(x)$.
Using the relation between $h_v(x)$ and $H_v(x)$
obtained from the Dirac equation, one can then
rewrite the QCD Lagrangian in powers of $1/m_Q$,
\begin{eqnarray}
        {\cal L} &=& \overline{h}_v iv \cdot D h_v + \overline{h}_v
                (i \not{\! \! D}_{\bot}) {1 \over 2m_Q + i v \cdot D
                -i\epsilon}(i \not{\! \! D}_{\bot}) h_v \nonumber \\
             &=& \overline{h}_v iv \cdot D h_v + \sum_{n=1}^{\infty}
                \Big({1 \over 2m_Q}\Big)^n \overline{h}_v (i \not{\!
                \! D}_{\bot}) (-i v \cdot D)^{n-1} (i \not{\! \!
                D}_{\bot}) h_v.  \label{ethql}
\end{eqnarray}
This is the effective Lagrangain for the heavy quark.
An equivalent derivation of the HQET via the QCD generating
functional can be found in Ref.\cite{Mannel}.
In the heavy mass limit ($m_Q\rightarrow\infty$),
only the first term in Eq.(\ref{ethql}) survives.
This leading order Lagragian is obviously spin and flavor
independent, which is the origin of the heavy quark
spin-flavor symmetry.
Note that, in Eq. (\ref{ethql}), the nonleading contributions contain
high order time derivatives.  Consequently the quantization
of the HQET beyond the leading order is rather cumbersome \cite{Lebed}.
As we will see later, this unpleasant feature does not exist in the
light-front formulation.

In order to gain a deeper understanding of heavy quark dynamics,
it is both necessary and important to study the symmetry breaking
effects caused by the higher order terms in the effective Lagrangian.
Similar terms can also arise in the $1/m_Q$ expansion of
heavy quark currents.
These higher order interactions would either spoil relations established
by HQS or introduce new transition form factors.
With HQET, one can in principle investigate these symmetry breaking
effects systematically.
However, in order to evaluate the various matrix elements involved,
one needs a detailed knowledge of the
structures of heavy hadron bound states.
To date, except for lattice simulation, a direct QCD approach
to the hadronic bound states does not exist,
and one has to rely on various phenomenological models, such
as the constitute quark model\cite{Isgur91}, the bag model\cite{bag}
and the QCD sum rules \cite{Neubert92}, to estimate these
matrix elements.
Since one does not know how to properly boost a constituent
quark bound state or a bag wave function to arbitrary velocities,
these models are, strickly speaking,
applicable only at the zero-recoil point.
However, to compare with experimental data,
matrix elements at various momentum transfer are required in general.

In the past few years, a boost-free relativistic approach to
hadronic bound states problem of QCD on the light-front
has attracted much attentions \cite{Wilson94,Perry,Zhang94}.   One of the
advantages for the light-front QCD approach is that the light-front
Hamiltonian field theory provides a direct way of calculating
relativistic bound states by solving Schr\"{o}dinger-type
eigenstate equations in a truncated Fock space \cite{Brodsky}.
It is well known that boost operations on the light-front are kinematic
and therefore it is easy to boost a hadron state
to any frame of reference when
its wave function is known in a particular Lorentz frame.
Moreover, special behavior of the light-front infrared singularity
may also lead to a possible understanding of the nontrivial QCD dynamics,
such as color confinement and
dynamical chiral symmetry breaking \cite{Wilson94}.
Nevertheless, light-front quantization for heavy quarks has only been
briefly explored in 1+1 dimensional model \cite{Burkart}.
Recently some light-front hadronic wave functions
have been constructed either phenomenologically \cite{BSW}
or from the light-front QCD sum rule \cite{Chern,Bely}.
They have been used quite
sucessfully in the calculations of the Isgur-Wise function and
other heavy hadron form factors.
Furthermore, inclusive heavy meson decays have also been
discussed on the light-front \cite{Jaffe}.

In order to better understand these light-front wave functions
and their applications in various heavy hadron processes,
we have recently reformulated the HQET
on the light-front \cite{ZLC}.
In the present paper, apart from providing a more detailed account of the
derivation of light-front heavy quark effective theory (LFHQET),
the construction of heavy meson bound states is also formulated.
We then derive the Isgur-Wise function using
the light-front wave functions so constructed;
the resulting expression is compatible with HQS
and valid for arbitrary recoil velocities.
The paper is organized as follows: in Sec. II,
LFHQET is derived, and its advantages over
the equal-time formulation are discussed.
In Sec. III, quantization procedure for the LFHQET is described.
In sec. IV, based on the LFHQET, heavy meson bound states are constructed
in the heavy mass limit.  An explicit calculation of the Isgur-Wise
function is given in Sec. V.
Finally, a summary is presented in Sec. VI.

%%%%%%%%%%%%%%%%%%%%%%%%%%%%%%%%%%%%%%%%%%%%%%%%%%%%%%%%%%%%%%%%%%%%%%%%%%%%%%%

\section{Light-front heavy quark effective theory}

In this section, the HQET is formulated on the light-front.
We shall use the following light-front notations:
The light-front coordinate is denoted by  $x^{\mu} = (x^+,x^-,x_{\bot})$
where $x^+ =x^0 + x^3$ is the
light-front time-like component, $x^- = x^0 - x^3$ and $x_{\bot}^i~
( i=1,2)$ the light-front longitudinal and transverse components
respectively.  With these notations, the product of two four-vectors
is given by $a \cdot b = {1\over 2}(a^+b^- + a^-b^+) - a_{\bot}
\cdot b_{\bot}$, and the light-front derivatives are written as
$\partial^- = 2 {\partial \over \partial x^+}$
(the light-front time derivative),
$\partial^+ = 2 {\partial \over \partial x^-}$,
and $\partial^i = {\partial \over \partial x_i}$
(the longitudinal and transverse derivatives respectively).
In order to express the final results in covariant forms,
we will also need the light-front unit vector $n^{\mu} = (0,1,0_{\bot})$,
such that the ``+" component of any four-vector $a$
can be written covariantly as $n \cdot a$.

In the conventional formulation of the HQET, the first step is
to separate the full heavy quark field $Q(x)$
into large and small components,
by means of the projection operators:
$\Lambda_\pm={1\over2}(1\pm\rlap\slash v)$.
The situation is somewhat different
in the framework of light-front field theory \cite{Zhang94}.
Here, before the $1/m_Q$ expansion is introduced,
the heavy quark field is first divided into two parts:
$Q(x) = Q_+(x) + Q_-(x)$,  with $Q_{\pm}(x) = \Lambda^{\pm} Q(x)
={1 \over 2} \gamma^0 \gamma^{\pm} Q(x)$.   The equation of motion for
$Q$ can then be rewritten as two coupled equations for $Q_\pm$:
\begin{eqnarray}
       i D^-Q_+(x) &=& ( i \alpha_{\bot} \cdot D_{\bot}
                + \beta m_Q) Q_- (x),   \label{lffd1}  \\
       i D^+Q_-(x) &=& ( i \alpha_{\bot} \cdot D_{\bot}
                + \beta m_Q) Q_+ (x),   \label{lffd2}
\end{eqnarray}
where $\alpha_{\bot} = \gamma^0 \gamma_{\bot}$ and $\beta =
\gamma^0$.  The above equations show that only the plus-component
$Q_+(x)$ is the dynamical field.  The equation of motion for the
minus-component $Q_-(x)$ does not contain
a light-front time derivative and therefore is a light-front constraint
that determines $Q_-(x)$ from $Q_+(x)$. In terms of $Q_+(x)$,
the heavy quark part of the QCD Lagrangian (\ref{1.1}) can be rewritten as

\begin{equation}
        {\cal L} = Q_+^{\dagger} i D^- Q_+ - Q_+^{\dagger} ( i
                \alpha_{\bot} \cdot D_{\bot} + \beta m_Q ) Q_- ,
        \label{2.3}
\end{equation}
where $Q_-$ can be eliminated by Eq.(\ref{lffd2}).

To derive the light-front HQET, we use the same
redefinition for the heavy quark field as in the equal-time case,
\begin{equation}
        Q(x) = e^{-i m_Q v \cdot x} {\cal Q}_v(x),  \label{nlfq1}
\end{equation}
but without imposing the separation of large and small components.
It follows that
\begin{equation}
        Q_+(x) = e^{-i m_Q v \cdot x} {\cal Q}_{v+}(x) ~~,~~~
        Q_-(x) = e^{-i m_Q v \cdot x} {\cal Q}_{v-}(x) . \label{nlfq2}
\end{equation}
Substituting these equations into Eq.(\ref{lffd2}), we obtain
\begin{equation}
        {\cal Q}_{v-} (x) = {1 \over m_Q v^+ + iD^+} \Big[i\alpha_{\bot}
                \cdot D_{\bot} + m_Q ( \alpha_{\bot} \cdot v_{\bot} +
                \beta) \Big] {\cal Q}_{v+} (x).
\end{equation}
It is worth noting that in the ordinary light-front formulation of
quantum field theory, the elimination of the dependent component
${Q_-}$ requires the choice of
the light-front gauge $A^+=0$, and a specification of the operator
$1/\partial^+$ which leads to severe light-front infrared problem
that has still not been completely understood\cite{Zhang93}.
However, for the heavy quark field with the redefinition of Eq.(\ref{nlfq1}),
the above problem does not occur since the elimination of the dependent
component ${{\cal Q}_{v-}}$ now depends on the operator $1/(m_Q
v^+ + iD^+)$ which has no infrared problem. Moreover, it has a well
defined series expansion in powers of $iD^+/m_Q$:
\begin{equation}
        {1 \over m_Q v^+ + iD^+} = \sum_{n=1}^{\infty} \Big( {1 \over
                m_Q v^+} \Big)^n ( -iD^+)^{n-1}.
\end{equation}
%Hereafter, the operator $1/(m_Q v^+ + iD^+)$ is always defined by
%the above expansion.
Thus,
\begin{eqnarray}
        {\cal Q}_{v-}(x) &=& \Big\{ { \alpha_{\bot} \cdot v_{\bot} + \beta
                \over v^+} + \sum_{n=1}^{\infty} \Big( {1 \over m_Q v^+}
                \Big)^{n}(-i D^+)^{n-1} ( i \vec{\alpha} \cdot \vec{D})
                \Big\} {\cal Q}_{v+}(x) \nonumber \\
        &=& \Big\{ { \alpha_{\bot} \cdot v_{\bot} + \beta \over v^+} +
                {1 \over m_Q v^+ + iD^+}(i \vec{\alpha} \cdot \vec{D})
                \Big\} {\cal Q}_{v+}(x),     \label{lfme}
\end{eqnarray}
where we have denoted
\begin{equation}
        \vec{\alpha} \cdot \vec{D}= \alpha_{\bot} \cdot D_{\bot} -
                { \alpha_{\bot} \cdot v_{\bot} + \beta \over v^+} D^+ .
\end{equation}

In the following, we show that
an alternative derivation based on the conventional way of eliminating
the dependent quark field component $Q_{-}$ gives the same result as above.
We shall work with the light-front gauge, in which $A^+=0$,
so that Eq. (\ref{lffd2}) becomes,
\begin{equation}
        Q_-(x) = {1 \over i \partial^+} ( i \alpha_{\bot} \cdot D_{\bot}
                + \beta m_Q) Q_+ (x).
\end{equation}
Using the integral definition\cite{Zhang93} of the operator $1/\partial^+$,
\begin{equation}
        \Big({1 \over \partial^+}\Big) f(x^-) = {1 \over 4}
                \int_{-\infty}^{\infty} dx'^- \varepsilon (x^- - x'^-)
                f(x'^-),
\end{equation}
where $\varepsilon(x) = -1,~ 0,~ 1$ for $x <0,~=0,~> 0$ respectively,
we have
\begin{equation}
   Q_-(x) = {1 \over i4} \int_{-\infty}^{\infty} dx'^- \varepsilon (x^-
           - x'^-) e^{-im_Q v \cdot \tilde x'} \Big[i \alpha_{\bot} \cdot
                D_{\bot} + m_Q (\alpha_{\bot} \cdot v_{\bot} + \beta)
                \Big] {\cal Q}_{v+}(\tilde x') , \label{dq1}
\end{equation}
where $\tilde x' \equiv (x^+, x'^-, x_\bot)$.
By repeated integration by parts, and
ignoring the surface terms (which are proportional to $\exp(-im_Q v
\cdot x)|_{x^- = \pm \infty}$, a highly oscillating term that can be
dropped), we finally find
\begin{equation}
        {\cal Q}_{v-} (x) = \Big\{ { \alpha_{\bot}
                \cdot v_{\bot} + \beta \over v^+} + \sum_{n=1}^{\infty}
                \Big( {1 \over m_Q v^+}\Big)^{n}  (-i \partial^+)^{n-1}
                ( i \vec{\alpha} \cdot \vec{D}) \Big\} {\cal Q}_{v+}(x),
\end{equation}
which is the same as Eq.(\ref{lfme}) in the light-front gauge.

Using Eq. (\ref{lfme}), one can rewrite the equation of motion for
${\cal Q}_{v+}(x)$, {\it i.e.}, Eq.(\ref{lffd1}), as:
\begin{equation}
        2(iv \cdot D) {\cal Q}_{v+} (x) =  (i \vec{\alpha} \cdot \vec{D})
                {v^+ \over m_Q v^+ + iD^+}(i \vec{\alpha} \cdot \vec{D})
                {\cal Q}_{v+}(x).  \label{lfhqem}
\end{equation}
Likewise, the heavy quark QCD Lagrangian (\ref{2.3}) can be reexpressed in
terms
of ${\cal Q}_{v+}$ alone. The complete $1/m_Q$ expansion is then given by
\begin{eqnarray}
        {\cal L}&=& {2 \over v^+} {\cal Q}_{v+}^{\dagger} (iv \cdot D)
                {\cal Q}_{v+} - {\cal Q}_{v+}^{\dagger}(i \vec{\alpha}
                \cdot \vec{D}){1 \over m_Q v^+ + iD^+}(i \vec{\alpha}
                \cdot \vec{D}){\cal Q}_{v+}(x). \nonumber \\
        &=& {2 \over v^+} {\cal Q}_{v+}^{\dagger} (iv \cdot D)
                {\cal Q}_{v+} - \sum_{n=1}^{\infty} \Big({ 1 \over m_Q v^+}
                \Big)^n {\cal Q}_{v+}^{\dagger} \Big\{(i\vec{\alpha}
                \cdot \vec{D}) (-i D^+)^{n-1} (i \vec{\alpha} \cdot
                \vec{D}) \Big\} {\cal Q}_{v+} (x) \nonumber \\
        &=& {\cal L}_0 + \sum_{n=1}^{\infty} {\cal L}_n . \label{lfhqetl}
\end{eqnarray}
This is the light-front effective heavy quark Lagrangian.
One can readily check that the equation of motion,
Eq.(\ref{lfhqem}), is consistent with this Lagrangian.
The dimensionless expansion
parameter in the above Lagrangian is indeed $\Lambda_{QCD}/m_Q$
as advertised earlier, since
the operator $(-iD^+)$ picks up the ``residual'' momentum of the heavy
quark, $k^+ = p^+ - m_Q v^+$, which is of the order $\Lambda_{QCD}$.

As mentioned earlier, in the above derivation of the light-front HQET,
unlike the equal-time case,
no constraint is imposed from the start to separate the large and small
components of the heavy quark field. In the present formalism, this
separation of the large and small components
is automatic.
To see this point more clearly, we rewrite the above results
in covariant forms. First let us define
\begin{equation}
        {\cal Q}_v = {\cal Q}_{v+} + {\cal Q}_{v-} \equiv h_v^L +
                H_v^L ,
\end{equation}
where $h_v^L$ is $m_Q$ independent and $H_v^L$ contains
all the $1/m_Q$ correction terms, {\it viz}.,
\begin{eqnarray}
        h_v^L &=& \Big\{ 1 + { \alpha_{\bot} \cdot v_{\bot} + \beta \over
                v^+} \Big\} {\cal Q}_{v+}, \label{llfhd1} \\
        H_v^L  &=& { 1 \over m_Q v^+ + iD^+}
                (i{\vec\alpha}\cdot{\vec D}) {\cal Q}_{v+} =
                - {\not{\! n} \over 2(m_Q
                n \cdot v + i n \cdot D)} (i\! \not{\! \! D}) h_v^L ,
\end{eqnarray}
where $n^{\mu}=(0,1,0_\bot)$ as defined earlier.
The superscript $L$ represents the fact that the large and small
components of the heavy quark field are
separated on the light-front. One can readily prove that the
zeroth order field operator $h_v^L$ has the desired property
\begin{equation}
        \not{\! v} h_v^L = h_v^L,
\end{equation}
whereas $H_v^L$ satisfies $\Lambda_+ H_v^L =0$.
Thus all $1/m_Q$ corrections are contained
in the light-front ``bad'' component $Q_-(x)$.
This fact provides a direct connection of the
$1/m_Q$ correction terms to high-twist operators,
as noticed in a QCD sum rule calculation of the Isgur-Wise
function \cite{Bely}. In terms of $h_v^L$, the covariant form of
the light-front effective heavy quark Lagrangian reads
\begin{eqnarray}
   {\cal L} &=& \overline{h}_v^L (i v \cdot D) h_v^L
            - \overline{h}_v^L (i\! \not{\! \! D}) { \not{\! n}
            \over 2(m_Q n\cdot v + i n \cdot D)} ( i\! \not{\! \! D})
            h_v^L \nonumber \\
   &=& \overline{h}_v^L (i v \cdot D) h_v^L
       - {1\over2}\sum_{l=1}^{\infty} \Big( {1 \over m_Q~n\cdot v} \Big)^l ~
       \overline{h}_v^L (i\! \not{\! \! D}) \not{\! n}
       (-in \cdot D )^{l-1}
       ( i\! \not{\! \! D}) h_v^L. \label{clfhql}
\end{eqnarray}

{}From Eq.(1.3), we
see that formally the light-front HQET has very similar
structure as the equal-time HQET. In the heavy mass limit,
the lowest order Lagrangian reads
\begin{equation}
        {\cal L}_0 = {2 \over v^+} {\cal Q}_{v+}^{\dagger} (i v \cdot D)
                {\cal Q}_{v+} = \overline{h}_v^L (i v \cdot D) h_v^L ,
\end{equation}
which is the same as the leading order equal-time effective theory,
clearly exhibits the familiar flavor and spin symmetries.
Note that spin symmetry on the light-front
is actually the same as helicity symmetry.

However, beyond the heavy mass limit, the LFHQET
has its advantages.
It is well known that, in the equal-time formulation,
the non-leading part of the HQET
contains high order time-derivatives.
This non-canonical structure of the HQET causes
certain difficulties in solving the theory \cite{Lebed}.
For instance, it is very difficult to write down
the  Hamiltonian to all orders in $1/m_Q$.
It is remarkable to see that, in the LFHQET, only the
linear time-derivative appears, and it resides
in ${\cal L}_0$ only.  The factor $ \not{\! n}$ in the
non-leading terms of the LFHQET eliminates all
light-front time derivative contributions,
as can be seen clearly from Eq.(\ref{lfhqetl}).  Therefore,
there is no difficulty in writing down the canonical conjugate field,
and hence the Hamiltonian from the effective Lagrangian
on the light-front.
Explicitly, the canonical conjugate of the dynamical variable
${\cal Q}_{v+}$ is
\begin{eqnarray}
        \Pi_{{\cal Q}_{v+}} = { \partial {\cal L} \over \partial
                (\partial^- {\cal Q}_{v+})} = i {\cal Q}_{v+}^{\dagger},
         \label{cong}
\end{eqnarray}
which does not involve any terms of order $1/m_Q$ or higher. The
light-front heavy-quark effective Hamiltonian density is then
given by:
\begin{eqnarray}
        {\cal H} &=& \Pi_{{\cal Q}_{v+}} \partial^- {\cal Q}_{v+}
                - {\cal L} \nonumber \\
                &=& { 1\over iv^+} {\cal Q}^{\dagger}_{v+} (v^-\partial^+
                -2v_{\bot} \cdot \partial_{\bot} ) {\cal Q}_{v+}
                -{2g \over v^+} {\cal Q}^{\dagger}_{v+} (v \cdot A)
                {\cal Q}_{v+} + {\cal H}_{m_Q}
\end{eqnarray}
with
\begin{equation}
        {\cal H}_{m_Q}= \sum_{n=1}^{\infty} {\cal H}_n
                = - \sum_{n=1}^{\infty} {\cal L}_n ,  \label{lfhqeh}
\end{equation}
and the light-front Hamiltonian is defined as
\begin{equation}
        H = P^- = \int dx^- d^2 x_{\bot} {\cal H} .
\end{equation}
This light-front heavy quark effective Hamiltonian can served as
a useful basis for constructing
the heavy hadron bound states. It is also interesting to note that the
light-front effective Hamiltonian ${\cal H}_n$ is precisely the
minus of the equal-time effective Lagrangian ${\cal L}_n$ given
by Eq.(\ref{lfhqetl}). This simple relation does not exist in
the equal-time HQET.  The reason is that, due to the existence of
high-order time-derivative in the equal-time HQET,
the effective Hamiltonian is minus of the effective
Lagrangian {\it plus} some noncanonical terms coming
from the unusual conjugate field.

This concludes the derivation of the LFHQET.
In order to apply this theory to practical problems,
one must first quantize it on the light-front.
We will turn to this subject in the next section.

%%%%%%%%%%%%%%%%%%%%%%%%%%%%%%%%%%%%%%%%%%%%%%%%%%%%%%%%%%%%%%%%%%%%%%%%%%%%%%%

\section{Light-front Quantization of HQET}

As shown earlier,
the equal-time heavy quark effective Lagrangian
contains higher order time derivatives, so that
it is very difficult to perform
a consistent canonical quantization beyond the limit
$m_Q\rightarrow \infty$ \cite{Lebed}.
However, as we have seen,
the light-front heavy quark effective Lagrangian only contains a linear
light-front time derivative term which resides in ${\cal L}_0$.
Thus the full light-front effective Lagrangian
can be easily quantized canonically.
By the light-front phase space quantization procedure \cite{Zhang93},
the basic anti-commutation relation is:
\begin{equation}
        \{ {\cal Q}_{v+}(x)~, ~ \Pi_{{\cal Q}_{v'+}}(y) \}_{x^+=y^+} =
                i\Lambda^+ \delta_{vv'}\delta(x^--y^-)\delta^2(x_\bot-y_\bot),
\label{lfhqqa}
\end{equation}
which is valid to all orders in $1/m_Q$.

In the limit $m_Q \rightarrow \infty$,
the light-front heavy quark field ${\cal Q}_+$
can be expanded in momentum space as
\begin{equation}
        {\cal Q}_{v+} (x) = \sum_{\lambda}
                \int{dk^+ d^2k_{\bot} \over
                2(2\pi)^3} \omega_{\lambda} b_v(k,\lambda)
                e^{-ikx} , \label{ex}
\end{equation}
where $k$ is the residual momentum of the heavy quark, $p=m_Qv+k$,
with $v \cdot k =0$ (mass-shell condition);
$\omega_{\lambda}$ is the plus-component of heavy quark spinor
which can be chosen to be momentum independent
in a particular representation of
the Dirac matrices \cite{Zhang93}, and it is normalized according to
$ \omega_{\lambda}^{\dagger} \omega_{\lambda'}
=\delta_{\lambda \lambda'}$, $\sum_{\lambda} \omega_{\lambda}
\omega^{\dagger}_{\lambda}=\Lambda^+$ (from Eq.(\ref{lfhqqa})).
$b_v(k,\lambda)$ is the heavy quark annihilation operator,
satisfying the basic anti-commutation relation
\begin{equation}
        \{b_v(k,\lambda) , b_{v'}^{\dagger}(k',\lambda') \} = 2(2\pi)^3
                \delta_{vv'} \delta(k^+-k'^+)\delta^2(k_\bot-k'_\bot)
                \delta_{\lambda \lambda'},
\end{equation}
where $\delta_{vv'}$ gives rise to the so-called
velocity superselection rule \cite{Georgi}.
Note that the antiquark part in Eq.(\ref{ex}) is dropped because
heavy quark-antiquark pair production is kinematically suppressed
at the scale we are interested in.

Feynman rules for the effective heavy quark field $Q_{v+}$ are
\begin{equation}
        \begin{array}{l} S_{{\cal Q}_{v+}} (k) = { i \over 2} {v^+
                \over v \cdot k},  \\
        \Gamma_{{\cal Q}_{v+}{\cal Q}'_{v+}g} = i{2 \over v^+} g
                T^a v^{\mu} \end{array}
\end{equation}
for the heavy quark propagator and the quark-gluon vertex respectively.

For practical calculations, it is sometimes more convenient to work with
the effective field $h_v^L(x)$, introduced in Section II,
since it represents the full leading order part
of the heavy quark field $Q(x)$.
The momentum space expansion of $h_v^L$ is given by
\begin{equation}
        h_v^L  (x) = \sum_{\lambda} \int {dk^+ d^2k_{\bot}
                \over 2(2\pi)^3} u(v,\lambda)b_v(k,\lambda) e^{-ikx} ,
\end{equation}
where the corresponding heavy quark spinor $u$ is defined as
\begin{equation}
        u(v,\lambda) = \Big\{ 1 + { \alpha_{\bot} \cdot v_{\bot} + \beta
                \over v^+} \Big\} \omega_{\lambda}  \label{3.6}
\end{equation}
and satisfies the following normalization conditions:
\begin{equation}
         \overline{u}(v,\lambda) u(v,\lambda') =
                {2 \over v^+}\delta_{\lambda\lambda'}~,
         ~~~~ \sum_{\lambda} u(v,\lambda)
               \overline{u}(v,\lambda) = {1 + \not{\! v}
               \over v^+}  .  \label{lfqfn}
\end{equation}

The corresponding Feynman rules for the $h^L_v$ field is given by,
\begin{equation}
        \begin{array}{l} S_{h^L_v} (k) = {i\over 2}{ 1+ \not{\! v}
                \over v\cdot k}, \\
        \Gamma_{h^L_v h^L_v g} = i{ g }T^a v^{\mu}.\end{array}
\end{equation}
This completes our discussion on the quantization of the LFHQET.

%%%%%%%%%%%%%%%%%%%%%%%%%%%%%%%%%%%%%%%%%%%%%%%%%%%%%%%%%%%%%%%%%%%%%%%%%%%%%%%

\section{Light-front Heavy Meson Bound States}

In this section, we outline the procedure for constructing
a heavy meson bound state wave function on the light front
\cite{Brodsky}.
In general, a hadronic bound states on the light-front can
be expanded in the Fock space composed of states with definite
number of particles.
Explicitly, a hadronic bound state with the total
longitudinal and transverse momenta $P^+, P_{\bot}$, and
helicity $\lambda$ can be written as
\def\pt{\tilde p}
\def\kt{\tilde k}
\def\dpt{d^3\tilde p}
\begin{equation}
        | \Psi(P^+, P_{\bot},\lambda) \rangle = \sum_{n,\lambda_i}
                \int \left( \prod_i {{\dpt_i} \over 2(2\pi)^3} \right)
                2 (2\pi)^3 \delta^3(\tilde P-\sum_i \pt_i)
                | n, \pt,\lambda_i \rangle
                \Phi_{n} (x_i,\kappa_{\bot i},\lambda_i), \label{lfwf}
\end{equation}
where $\tilde p\equiv (p^+, p_\bot)$, so that $\dpt = dp^+dp^2_{\bot}$,
and $\delta^3(\pt-\pt')=\delta(p^+-p'_+)\delta^2(p_\bot-p'_\bot)$;
$|n, \pt, \lambda_i \rangle$, is the Fock state consisting of $n$
constituents, each of which carries momentum $\pt_i$ and helicity
$\lambda_i$ ($\sum_i \lambda_i = \lambda$);
$\Phi(x_i,\kappa_{\bot i},\lambda_i)$ is the corresponding amplitude
which depends on $\lambda_i$, the longitudinal momentum fraction
$x_i$, and the relative transverse momentum $\kappa_{\bot i}$:
\begin{equation}
x_i = { p_i^+ \over P^+}~~, ~~~ \kappa_{i\bot} = p_{i\bot} - x_i P_{\bot}.
\end{equation}

The eigenstate equation that the wave functions obey on the light-front
is obtained from the operator Einstein equation $P^2 =P^+P^- - P_{\bot}^2
= M^2$:
\begin{equation}
        H_{LF} | P^+, P_{\bot},\lambda \rangle  = { P_{\bot}^2
                + M^2 \over P^+ } | P^+, P_{\bot},
                \lambda \rangle
\end{equation}
where $H_{LF}={P}^-$ is the light-front Hamiltonian.
Explicitly, for a meson wave function, the
corresponding light-front bound state equation is:
\begin{equation}
        \Big(M^2 - \sum_i { \kappa_{i\bot}^2 + m_i^2 \over x_i} \Big)
                \left[\begin{array}{c} \Phi_{q\bar{q}} \\
                \Phi_{q\bar{q}g} \\ \vdots \end{array} \right]
                  = \left[ \begin{array}{ccc} \langle q \bar{q}
                | H_{int} | q \bar{q} \rangle & \langle q \bar{q} | H_{int}
                | q \bar{q} g \rangle & \cdots \\ \langle q \bar{q} g
                | H_{int} | q \bar{q} \rangle & \cdots & ~~  \\ \vdots &
                \ddots & ~~ \end{array} \right] \left[\begin{array}{c}
                \Phi_{q\bar{q}} \\ \Phi_{q\bar{q}g} \\ \vdots \end{array}
                \right], \label{lfbe}
\end{equation}
where $H_{int}$ is the interaction part of $P^-$.

Obviously to solve the above equation from QCD with the whole
Fock space is impossible.  Nevertheless HQS can still
bring great simplification to the problem.
First of all we note that, on the light-front, the total
helicity of a heavy meson is simply the sum of the helicity
of the heavy quark and the
total helicity of the light quark sector (the so-called brown
muck which carries total spin 1/2;
the brown muck of a baryon is more complex, and will
not be discussed here.)
HQS implies that, in the limit $m_Q \rightarrow \infty$,
the spin of the heavy quark is
decoupled from that of the light quark part,
because the heavy quark interacts with
the light quark part only through
spin-independent soft gluon exchanges.
Thus, for a heavy meson, we can approximate
the general expression of the light-front bound states
Eq.(\ref{lfwf}) as follows:
\begin{equation}
        | \Psi (P^+, P_{\bot}, \lambda) \rangle =
                  \sum_{\lambda_Q\lambda_q}
        \int {\dpt_Q\dpt_q\over2(2\pi)^3 }\delta^3(\tilde P-\pt_Q-\pt_q)
                \Phi_{Q\bar{q}}(x,\kappa_{\bot},\lambda_Q,\lambda_q)
                | Q(p_Q,\lambda_Q), \bar{q}(p_q,\lambda_q) \rangle,
                \label{lfhqwf}
\end{equation}
where $P=Mv$, $M$ is the mass of meson and $v^{\mu}$ is its four velocity;
while
\begin{equation}
                | Q(p_Q,\lambda_Q), \bar{q}(p_q,\lambda_q) \rangle
                =b_Q^\dagger(p_Q,\lambda_Q)
                 d_q^\dagger(p_q,\lambda_q) |0\rangle,
\end{equation}
and $d^\dagger_{q}$ should be regarded as the creation operator of a
constituent light antiquark (brown muck), consisting of the valence
current anti-quark and a sea of gluon and quark-antiquark pairs.
Consequently, contribution from the higher Fock states
may be replaced by an effective two-body interaction kernel,
so that Eq.(\ref{lfbe}) is reduced to a light-front
Bethe-Salpeter equation:
\begin{equation}
        \Big( M^2 - M_0^2 \Big) \Phi_{Q \bar{q}} (x,
           \kappa_{\bot}) = \int {dx' d^2 \kappa'_{\bot} \over 2 (2\pi)^3}
           V_{eff} (x,\kappa_{\bot},x', \kappa'_{\bot}) \Phi_{Q \bar{q}}(x',
                \kappa'_{\bot}),  \label{lfbse}
\end{equation}
and
\begin{equation}
        M_0^2 = {\kappa_{\bot}^2 + m_q^2 \over x} + {\kappa_{\bot}^2 +m_Q^2
                \over 1-x} .
\end{equation}
In principle, the two-body effective interaction kernel $V_{eff}$
should be derived from the leading order light-front
heavy quark effetive Hamiltonian,
plus the full QCD Hamiltonian for the light quarks at
the hadronic scale.
As is well known, the latter is very complicated even
in the naive canonical case \cite{Zhang93},
and to derive $V_{eff}$ is beyond the scope of this paper.
We will leave this subject for future investigation.
Until a way is found to solve the light-front bound state dynamics,
we would have to be contented with a phenomenological amplitude
for $\Phi_{Q\bar{q}}$.
One example that has been often used in the literature is
the so-called BSW amplitude \cite{BSW},
\begin{equation}
        \Phi_{BSW}(x,k_{\bot}) = {\cal N} \sqrt{x(1-x)}
                \exp \Big( -{\kappa_{\bot}^2 \over 2\omega^2} \Big)
                \exp \Big( - {M^2 \over 2\omega^2} (x -x_0)^2 \Big) ,
                 \label{bsw}
\end{equation}
where ${\cal N}$ is the normalization constant,
$x$ is the longitudinal momentum fraction carried by the light quark,
$x_0=\left({1\over2}-{m_Q^2-m_q^2\over 2M^2}\right)$, and
$\omega$ is a parameter related to the physical size of the meson.
Other forms, such as the Gaussian type \cite{GI,GWI},
are also possible, but we shall not dwell on this matter further.

Spin is always a troublesome issue in the light-front approach.
For example, the heavy meson light-front bound state we have
constructed is labeled by helicity rather than spin.
However for practical applications
physical states with definite spins are needed.
This discrepancy is usually remedied by introducing
the so-called Melosh rotation \cite{Melosh},
which transforms a single particle state from the
light-front helicity basis to the ordinary spin basis,
\begin{equation}
        R(x_i,k_{\bot},m_i) =  { m_i + x_i M_0 - i \sigma \cdot
                ({\bf n} \times \kappa_{\bot} ) \over \sqrt{ (m_i +
                x_i M_0)^2 + \kappa_{\bot}^2}} , \label{4.10}
\end{equation}
where ${\bf n} = (0, 0, 1)$.
With the Melosh transformation incorporated, the light-front heavy
meson bound state with a definite spin can be expressed as
follows \cite{Ter}
\begin{eqnarray}
        | \Psi (P^+, P_{\bot},S,S_z) \rangle &=& \sum_{\lambda_Q\lambda_q}
        \int {\dpt_Q\dpt_q\over2(2\pi)^3 }\delta^3(\tilde P-\pt_Q-\pt_q)
                \Phi_{Q\bar{q}}(x,\kappa_{\bot}) \nonumber \\
                &&\qquad\qquad \times R^{SS_z}_{\lambda_Q \lambda_q}
                (x,\kappa_{\bot})| Q(p_Q,\lambda_Q),
                \bar{q}(p_q,\lambda_q)\rangle,
                                                  \label{lfhqwf1}
\end{eqnarray}
where
\begin{equation}
        R_{\lambda_Q \lambda_q}^{SS_z} (x,\kappa_{\bot}) = \sum_{s_1s_2}
                 \langle \lambda_Q | R^{\dagger} (1-x, \kappa_{\bot}, m_Q)
                 | s_1 \rangle \langle \lambda_q | R^{\dagger} (x,
                 - \kappa_{\bot}, m_q) | s_2 \rangle \langle {1\over 2}
                 s_1 {1\over 2}s_2 | SS_z \rangle,   \label{mttb}
\end{equation}
and $\langle {1\over 2}s_1 {1\over 2}s_2 | SS_z \rangle$ is the
Clebsch-Gordon coefficient. A covariant form of Eq.(\ref{mttb}) has
been derived by Jaus \cite{Jaus}, which makes practical
calculations very convenient:
\begin{equation}
        R_{\lambda_Q \lambda_q}^{SS_z} (x,\kappa_{\bot}) = \sqrt{ n\cdot
                p_Q n \cdot p_q \over 2[M_0^2 - (m_Q-m_q)^2] }
                \overline{u}(p_Q, \lambda_Q) \Gamma v(p_q,\lambda_q),
\end{equation}
where
\begin{eqnarray}
        \Gamma &=& \gamma^5 ~~~~~  ({\rm for~pseudoscalar}, S = 0), \\
        \Gamma &=& - \not{\! \epsilon}(S_z) + {\epsilon \cdot (p_Q-p_q)
           \over M_0 + m_Q + m_q} ~~~~~ ({\rm for~vector}, S=1)
\end{eqnarray}
with
\begin{eqnarray}
       \epsilon^{\mu}(\pm 1) &=& \Big({2 \over P^+} \epsilon_{\bot}
                P_{\bot}, 0, \epsilon_{\bot} \Big)~,~~ \epsilon_{\bot}
                (\pm 1) = \mp (1, \pm i)/\sqrt{2} \nonumber \\
         \epsilon^{\mu}(0) &=& - {1 \over M_0} \Big( {-M_0^2 + P_{\bot}^2
                 \over P^+}, P^+ , P_{\bot} \Big),
\end{eqnarray}
and the spinor $u(p,\lambda)$ has the same form as Eq.(\ref{3.6}).
Eq.(\ref{lfhqwf1}) is the phenomenological light-front heavy meson bound
state that has been widely used in the study of heavy hadronic
dynamics \cite{BSW,NeuRie}.

However the heavy meson bound state so constructed still explicitly
depends on the heavy quark mass $m_Q$, and so is inconvenient
from the view point of HQET.
To calculate heavy hadron matrix elements,
we would like to use wave functions constructed in the heavy mass limit,
and then $1/m_Q$ corrections can be
treated order by order within the framework of LFHQET.
{}From Eq.(\ref{lfhqwf1}), a heavy meson bound state
in the heavy quark limit is given by
\begin{equation}
        | \Psi (v, S, S_z) \rangle = \sum_{\lambda_Q \lambda_q}
           \int {d^3 \kt  \dpt_q \over 2 (2\pi)^3 }
           \delta^3(\Lambda_Q\tilde v - \kt -\pt_q) \Phi_{Q\bar{q}}
         (x,\kappa_{\bot}) R_{\lambda_Q \lambda_q}^{SS_z}
         | Q_v(k,\lambda_Q), \bar{q}(p_q, \lambda_q) \rangle ,
         \label{4.13}
\end{equation}
where $\Lambda_Q=M-m_Q, x={p_q^+/(Mv^+)}, \kappa_\bot=p_{q\bot}-x(Mv_\bot)$,
and the Melosh transformation matrix element is reduced to
\begin{equation}
        R_{\lambda_Q \lambda_q}^{00} = \sqrt{ v^+
                p^+_q \over 4(\Lambda_Q+m_q) }~ \overline{u}(v,
                \lambda_Q) \gamma^5 v (p_q,\lambda_q)
        \label{4.14}
\end{equation}
for a pseudoscalar meson, and
\begin{equation}
        R_{\lambda_Q \lambda_q}^{1S_z} = - \sqrt{ v^+
                p_q^+ \over 4(\Lambda_Q+m_q) }~ \overline{u}(v,
                \lambda_Q) \! \not{\! {\epsilon}}(S_z)
                v (p_q,\lambda_q),  \label{4.19}
\end{equation}
for the vector meson; and the polarization vector becomes
\begin{equation}
        {\epsilon}^{\mu} (\pm1) = \Big( {2 \over v^+}
         \epsilon_{\bot} \cdot v_{\bot}, 0, \epsilon_{\bot} \Big)~,~~
        {\epsilon}(0) = - \Big( {v^2_{\bot} -1 \over v^+}, v^+,
                v_{\bot} \Big).     \label{4.20}
\end{equation}
where we have approximately let $p_q=(M-m_Q)v=\Lambda_Q v$
in the Melosh transformation matrix elements.
This is because in the symmetry limit the heavy quark spinor
in the Melosh transformation matrix element is independent
of the residual momentum $k$ (or the relative momentum $x, k_{\bot}$),
as can be seen from Eqs. (\ref{4.14}-\ref{4.19}).
Thus the residual momentum dependence in the light quark
spinor should also be very weak in order that the light-front
heavy meson state carries a fixed spin.
The normalization condition for the state $|\Psi(v,S,S_z)\rangle$
is taken to be
\begin{equation}
        \langle \Psi(v',S',S'_z) | \Psi(v,S,S_z) \rangle = 2(2\pi)^3
                P^+ \delta^3(\tilde v-\tilde v')
                \delta_{S'S}\delta_{S'_zS_z}, \label{4.17}
\end{equation}
which leads to
\begin{equation}
       \int {dx d^2 \kappa_{\bot} \over 2 (2\pi)^3}
                |\Phi_{Q\bar{q}}(x,\kappa_{\bot})|^2 = 1. \label{4.18}
\end{equation}

Thus we have constructed a light-front heavy meson bound state
in the symmetry limit ($m_Q \rightarrow \infty$) which
has definite spin and parity.
In the next section, we shall derive the Isgur-Wise
function from this light-front wave function.

%%%%%%%%%%%%%%%%%%%%%%%%%%%%%%%%%%%%%%%%%%%%%%%%%%%%%%%%%%%%%%%%%%%%%%%%%%%%%%

\section{Isgur-Wise Function}

In the LFHQET, as in the equal-time formulation,
one can readily show that there exists an universal function
describing the weak transitions between heavy mesons.
To do so, we first expand the weak
heavy quark current in $1/m_Q$ on the light-front, namely,
\begin{eqnarray}
    \overline{Q}^j(x)\Gamma Q^i(x)&=&e^{i(m_{Q^j}v'-m_{Q^i}v)\cdot  x }
        {\cal Q}_{v'+}^{j\dagger} \left(1 + { \alpha_{\bot} \cdot v'_{\bot}
        +\beta \over v'^+} + (-i \vec\alpha \cdot
        \stackrel{\leftarrow}{D}_{\bot} ) \sum_{n=1}^{\infty}
        \Big( {1 \over m_Qv'^+} \Big)^n (i
        \stackrel{\leftarrow}{D}^+)^{n-1} \right) \nonumber\\
   & & ~~ \times  \gamma^0 \Gamma \left(1+{\alpha_{\bot} \cdot v_{\bot}+
        \beta \over v^+} + \sum_{n=1}^{\infty} \Big( {1 \over m_Qv^+}
        \Big)^n (-iD^+)^{n-1} (i \vec\alpha \cdot \vec D_{\bot} ) \right) {\cal
        Q}^i_{v+}(x),  \label{lfcl}
\end{eqnarray}
where $\Gamma$ stands for an arbitrary Dirac matrix
($\gamma_5$, $\gamma_\mu$, {\it etc}.).
In the heavy mass limit, it reduces to the following familiar from:
\begin{equation}
        \overline{Q}^j(x)\Gamma Q^i(x) \rightarrow
                e^{i(m_{Q^j}v'-m_{Q^i}v) \cdot x }
                \overline{h}_{v}^{jL}(x) \Gamma h_v^{iL} (x) ,
\end{equation}
which shows that, apart from a trivial exponential factor,
the effective current does
not depend on the heavy quark masses, and hence is flavor independent.
Consequences of the spin and flavor symmetries can be readily derived
using this zeroth order heavy quark current.
Consider the following matrix elements, for example,
\begin{equation}
        \langle P_{Q^j} (v') | \overline{h}_{v'}^{jL} \Gamma
                h_v^{iL} | P_{Q^i} (v) \rangle
                 ~~ {\rm and} ~~ \langle P^*_{Q^j} (v') |
                \overline{h}_{v'}^{jL} \Gamma h_v^{iL} | P_{Q^i} (v)
                \rangle,            \label{5.3}
\end{equation}
where $P_Q$ and $P^*_Q$ represent respectively
a pseudoscalar meson and a vector meson containing
a single heavy quark $Q$.
Formally the heavy mesons states can be represented by the
interpolating fields: $| P_{Q^i} (v)\rangle = \sqrt{M_i}~
\overline{h}_v^{iL}\gamma_5 \ell_v | 0 \rangle$,
$| P^*_{Q^i} (v) \rangle = \sqrt{M^*_i}~
\overline{h}_v^{iL}\not{\!\epsilon}~\ell_v | 0 \rangle$,
where the mass factors are introduced for normalization purpose only, and
$\ell_v$ stands for the fully interacting light anti-quark (or brown muck)
inside a heavy meson moving with velocity $v$.
$\ell_v$ carries the quantum numbers of the valence light anti-quark,
but is independent of
the spin and flavor of the associated heavy quark.
As we have seen in Section III, the propagator for the
$h^L_v$ field is proportional to $(1+\rlap\slash v)/2$.
It is then easy to show that, in the symmetry limit,
the heavy meson transition matrix elements take the
familiar forms \cite{Wise2}
\begin{eqnarray}
        & & \langle P_{Q^j} (v') | \overline{h}_{v'}^{jL} \Gamma
                h_v^{iL} | P_{Q^i} (v) \rangle
                = \sqrt{M_i M_j}
                Tr\Big\{ \gamma_5 \Big( {1+ \not{\! v}'
                \over 2} \Big)\Gamma \Big( {1+\not{\! v} \over 2} \Big)
                \gamma_5 {\cal M} \Big\} \label{5.4}\\
        & & \langle P^*_{Q^j} (v') |
                \overline{h}_{v'}^{jL} \Gamma h_v^{iL} | P_{Q^i} (v)
                \rangle = \sqrt{M_i M^*_j}
                Tr\Big\{ \not{\! \epsilon}^* \Big( {1 + \not{\! v}'
                \over 2} \Big) \Gamma \Big( {1+ \not{\! v} \over 2}
                \Big) \gamma_5  {\cal M} \Big\}.   \label{5.5}
\end{eqnarray}
where ${\cal M}$ is the transition matrix element
for the light anti-quark (brown muck):
\begin{equation}
        {\cal M} = \langle 0 | \overline{\ell}_{v'} \ell_v | 0 \rangle
                \rightarrow  \xi (v \cdot v') I.
\end{equation}
Thus HQS implies that the transition matrix elements
(\ref{5.3}) are described by a single form factor $\xi(v \cdot v')$,
known as the Isgur-Wise function.

Next, we explicitly derive Eq. (\ref{5.4}-\ref{5.5}) from the
light-front bound state wave functions of the general form (\ref{4.13}),
and thereby extract the Isgur-Wise function in terms of the light-front
amplitudes.
The hadronic matrix element for $B$ to $D$ transition is given by
\begin{eqnarray}
      & &  \langle D(v',0,0) |
    \overline{h}_{v'}^{cL}\Gamma h_v^{bL} | B(v,0,0)
    \rangle \nonumber\\
      & &\qquad=\int {\dpt_q\dpt'_q \over [2(2\pi)^3]^2}
          \Phi_D^*(x',\kappa'_\bot)\Phi_B(x,\kappa_\bot)
          R^{\dagger 00}_{\lambda_c\lambda'_q}
          R^{00}_{\lambda_b\lambda_q} \nonumber\\
    & &\qquad\qquad\times \langle c_{v'}
           (\Lambda_c v - p'_q,\lambda_c) |
           \overline{h}_{v'}^{cL}\Gamma h_v^{bL} |
           b_v(\Lambda_b v - p_q,\lambda_b)\rangle
           \langle \bar{q}(p'_q,\lambda'_q)|
           \bar{q}(p_q,\lambda_q)\rangle.
\label{5.7}
\end{eqnarray}
Since $\Lambda_b=\Lambda_c$ in the heavy quark limit, and
\begin{eqnarray}
&& \langle \bar{q}(p'_q,\lambda'_q) |
   \bar{q}(p_q,\lambda_q)\rangle
=2 (2\pi)^3\delta^3(\pt_q-\pt'_q)\delta_{\lambda_q\lambda'_q},\\
&&
\langle c_{v'}(\Lambda_c v- p'_q,\lambda_c) |
\overline{h}_{v'}^{cL}\Gamma h_v^{bL} |
b_v(\Lambda_b v-p_q,\lambda_b)\rangle
= \overline u(v',\lambda_c) \Gamma u(v,\lambda_b),
\end{eqnarray}
making use of relation (\ref{lfqfn}), we obtain
\begin{eqnarray}
        & &\langle D(v',0,0) | \overline{h}_{v'}^{cL}
                \Gamma h_v^{bL} | B(v,0,0) \rangle \nonumber\\
        & &\qquad=\sqrt{M_B M_D}~
            \zeta(v,v')~ Tr \Big\{ \gamma_5 \Big({1 + \not{\! v}'
                \over 2}\Big) \Gamma \Big({1 + \not{\! v} \over
                2}\Big) \gamma^5) \Big\};
\end{eqnarray}
similarly for $B$ to $D^*$ transition, we have
\begin{eqnarray}
       && \langle D^*(v',S,S_z) | \overline{h}_{v'}^{cL}
                \Gamma h_v^{bL} | B(v,0,0) \rangle \nonumber\\
       &&\qquad = \sqrt{M_B M_{D^*}}~\zeta(v,v')~
                Tr \Big\{ \not{\!\epsilon^*} \Big({1 + \not{\! v}'
                \over v^+}\Big) \Gamma \Big({1 + \not{\! v} \over
                v^+}\Big) \gamma^5) \Big\} ,
\end{eqnarray}
where $\epsilon$ is given by Eq. (\ref{4.20}).
The universal Isgur-Wise function
appearing in the above expressions is given by
\begin{equation}
   \zeta(v,v') = \sqrt{{M_B\over M_D}z}
   \int {dx d^2\kappa_\bot \over 2 (2\pi)^3}
   \Phi_D^*(x',\kappa'_{\bot}) \Phi_B(x,\kappa_{\bot}), \label{5.12}
\end{equation}
where $z \equiv {v^+/ v'^+}$, $x'={M_B\over M_D}zx$,
and $\kappa'_\bot=\kappa_\bot+xM_B(v_\bot-zv'_\bot)$.
To see the covariant structure of $\zeta(v,v')$,
without the loss of generality,
we can choose a frame where $v_\bot=v'_\bot=0$;
this is the most natural choice for light-front calculations.
In such a frame $\zeta$ is a function of $z$ only, and
$z$ can be expressed in terms of $v\cdot v'$ as
\begin{equation}
       z_{\pm} = v \cdot v' \pm \sqrt{(v \cdot v')^2 -1} .\label{z}
\end{equation}
where the +($-$) sign corresponds to $v^3$ greater(less) than $v'^3$,
and $z_+=1/z_-$.
In the rest frame of the $B$ meson,
this sign ambiguity corresponds to whether one chooses the velocity
of the $D$($D^*$) meson, $\vec v~'$,
to be in the negative or positive z-direction.
Since physically these two situations are indistinguishable,
we must have
\begin{equation}
      \zeta(z)=\zeta(1/z)=\xi(v\cdot v'),\label{zeta}
\end{equation}
which puts a constraint on the light-front amplitudes.
Furthermore, it is interesting to note that this
constraint condition can also be derived
by demanding that altering the order of the integrations
in Eq. (\ref{5.7}) does not change the final result.

In the symmetry limit, the Isgur-Wise function, Eq. (\ref{5.12}),
should be independent of all heavy meson masses.
This property can be explicitly checked
by observing that, when $m_Q \rightarrow \infty$,
the light-front amplitude
must have the following scaling behavior,
\begin{equation}
\Phi_{Q\bar{q}}(x,\kappa_\bot)\rightarrow
\sqrt{M}~\tilde\Phi(M x, \kappa_\bot),
\label{5.15}
\end{equation}
where the factor $\sqrt{M}$ ($M$ being the meson mass)
comes from the particular normalization
we have assumed for the physical state in Eq. (\ref{4.17}).
The reason why the light-front heavy meson wave function should have
such asymptotic form is as follows.
Since $x$ is the longitudinal momentum fraction
carried by the light quark,
hence the meson wave function should be sharply peaked near
$x\sim\Lambda_{QCD}/M$.
It is then clear that only terms of the form ``$Mx$" survive
in the wave function as $M(m_Q)\rightarrow\infty$
\footnote{Note that $Mx=p^+_q$ in the rest frame of the heavy meson.}.
With Eq. (\ref{5.15}), Eq. (\ref{5.12}) can be rewritten as
\begin{equation}
   \zeta(z) = \sqrt{z}\int_0^\infty {dX}
              \int{d^2 \kappa_{\bot} \over 2 (2\pi)^3}
           \tilde\Phi^*(X',\kappa_{\bot}) \tilde\Phi(X,\kappa_{\bot}),
                     \label{IWF}
\end{equation}
where $X\equiv M_B x$, $X'\equiv M_D x'$, and $X'= Xz$.
Now it is evident that the Isgur-Wise function $\zeta(z)$,
or $\xi(v\cdot v')$, is totally
independent of the heavy meson masses, not even their ratio \cite{NeuRie}.
Furthermore we also see that, at the zero-recoil point ($v\cdot v'=1$),
Eq. (\ref{IWF}) reduces to the normalization condition (\ref{4.18})
in the symmetry limit; hence $\xi(1)=\zeta(1)=1$ as required.

In other works which also use light-front wave functions,
hadronic from factors are usually evaluated either at the
maximum recoil point $(P-P')^2=0$, or for $(P-P')^2 \le 0$,
and special techniques are required to cover the whole
kinematic region of interest \cite{Jaus,NeuRie}.
This is not the case here.
In this paper, the Isgur-Wise function is derived without
assuming a particular value for $(P-P')^2$.
Hence Eq. (\ref{IWF}) is quite general,
and valid for arbitrary momentum transfers.

In the following, we explicitly calculate the Isgur-Wise function
for model light-front amplitudes.
In the heavy quark limit, one can easily show that
the phenomenological BSW wave function given in Eq. (\ref{bsw})
does have the correct asymptotic form (\ref{5.15}), with
\begin{equation}
\tilde\Phi_{BSW}(x,\kappa_\bot)=
\sqrt{32}\left({\pi\over\omega^2}\right)\sqrt{Mx}~
{\exp}\left({-\kappa_\bot^2\over 2\omega^2}\right)
{\exp}\left({-M^2x^2\over 2\omega^2}\right).
\label{5.17}
\end{equation}
Combining this expression and Eq. (\ref{IWF}), we find
\begin{equation}
\zeta(z)={2z\over 1+z^2},
\end{equation}
which indeed satisfies the consistency condition (\ref{zeta}).
With relation (\ref{z}), the Isgur-Wise function in the symmetry limit
can be expressed in terms of $v\cdot v'$, {\it viz.},
\begin{equation}
\xi(v\cdot v')={1\over v\cdot v'}.
\end{equation}
One can also check that the slope of $\xi(v\cdot v')$
at the zero-recoil point ($v\cdot v'=1$),
\begin{equation}
\rho^2 \equiv\ - \xi'(1) = 1,
\end{equation}
satisfies the Bjorken constraint of $\rho^2 > 1/4$ \cite{Bjorken}.
Moreover, it is in excellent agreement with the recent experimental
result from CLEO $\rho^2 = 1.01 \pm 0.15 \pm 0.09$ \cite{CLEO}.

%%%%%%%%%%%%%%%%%%%%%%%%%%%%%%%%%%%%%%%%%%%%%%%%%%%%%%%%%%%%%%%%%%%%%%%%%%%%%%%

\section{Summary}

To summarize, in this paper, we have explored in details the HQET
and the $1/m_Q$ expansion on the light-front. In the heavy quark
mass limit, the light-front formulation reproduces the heavy quark
spin-flavor symmetry, as in the equal time case.
However, the structure of the LFHQET is rather simple, so that
canonical quantization present no difficulty, and the Hamiltonian
is well defined to all the orders in $1/m_Q$, which is
in contrast to the equal time approach where since the non-leading
terms contain high order time derivatives,
the canonical procedures are not valid for
quantizing the theory and constructing the Hamiltonian.
In Section IV, we construct the light-front heavy meson
bound states in the $m_Q \rightarrow \infty$ limit for performing
practical evaluation of heavy hadron dynamics within LFHQET.
Finally, Isgur-Wise function is derived from the light-front
heavy meson wave functions, and the result is a general expression
valid for arbitrary recoil velocities. For the asymptotic form of
the BSW amplitude in the $m_Q \rightarrow \infty$ limit, we find that
the Isgur-Wise function $\xi(v \cdot v') = 1/v \cdot v'$ and
its slope at the zero-recoil point is $\rho^2 = - \xi'(1) = 1$
which is in excellent agreement with the recent CLEO result
of $\rho^2 = 1.01 \pm 0.15 \pm 0.09$.

%%%%%%%%%%%%%%%%%%%%%%%%%%%%%%%%%%%%%%%%%%%%%%%%%%%%%%%%%%%%%%%%%%%%%%%%%%%%%%%

\acknowledgements

This work was supported in part by the National Science Council
of the Republic of China under grant numbers NSC84-2112-M-001-036,
NSC84-2816-M-001-012L, and NSC84-2112-M-009-024.

%%%%%%%%%%%%%%%%%%%%%%%%%%%%%%%%%%%%%%%%%%%%%%%%%%%%%%%%%%%%%%%%%%%%%%%%%%%%%%%

\vspace{0.5in}

\noindent E-mail addresses:

phcheung@ccvax.sinica.edu.tw

wzhang@phys.sinica.edu.tw

glin@beauty.iop.nctu.edu.tw

%%%%%%%%%%%%%%%%%%%%%%%%%%%%%%%%%%%%%%%%%%%%%%%%%%%%%%%%%%%%%%%%%%%%%%%%%%%%%%%

% References

\end{document}